\documentclass[preprint]{aastex}

\usepackage{color}
\defcitealias{kim07}{KK07}
\newcommand{\kk}{\citetalias{kim07}}
\newcommand{\oops}[1]{#1}
\newcommand{\vct}[1]{\mathbf{#1}}
\newcommand{\cs}{c_{\rm s}}
\newcommand{\mach}{\mathcal{M}_p}
\newcommand{\pert}{\mathcal{N}}
\newcommand{\amin}{\alpha_{\min}}
\newcommand{\degree}{\ensuremath{^\circ}}

\shorttitle{GRAVITATIONALLY INDUCED DENSITY WAKE}
\shortauthors{H.\ KIM}

\begin{document}
\title{Gravitationally Induced Density Wake of a Circularly Orbiting Object
As an Interpretative Framework of Ubiquitous Spirals and Arcs}
\author{Hyosun Kim\altaffilmark{1}}
\altaffiltext{1}{Academia Sinica Institute of Astronomy and Astrophysics, 
  P.O. Box 23-141, Taipei 10617, Taiwan; hkim@asiaa.sinica.edu.tw}

\begin{abstract}
An orbiting object in a gas rich environment creates a gravitational 
density wake containing information about the object and its orbit. 
\oops{Using linear perturbation theory,} we analyze the observable 
properties of the gravitational wake due to the object circularly 
moving in a static homogeneous gaseous medium, in order to derive 
the Bondi accretion radius $r_B$, the orbital distance $r_p$, and 
the Mach number $\mach$ of the object. Supersonic motion, producing 
a wake of spiral-onion shell structure, exhibits a single-armed 
Archimedes spiral \oops{and two-centered circular arcs with respect 
to the line of sight}. The pitch angle, arm width, and spacing of 
the spiral pattern are entirely determined by the orbital distance 
$r_p$ and Mach number $\mach$ of the object. The arm-interarm density 
contrast is proportional to $r_B$, decreasing as a function of distance 
with a power index of $-1$. \oops{The background density distribution 
is globally changed from initially uniform to centrally concentrated.}
The vertical \oops{structure} of the wake is 
manifested as circular arcs with the center at the object location. 
The angular extent of the arcs is determined by the Mach number 
$\mach$ of the object motion. \oops{Diagnostic probes of nonlinear
wakes such as} a detached bow shock, an absence of the definite 
inner arm boundary, the presence of turbulent low density eddies, 
and elongated shapes of arcs are \oops{explained in the extension 
of the linear analysis}. The density 
enhancement at the center is always $r_B/r_p$ independent of the 
nonlinearity, suggesting that massive objects can substantially 
modify the background distribution.
\end{abstract}


\section{INTRODUCTION}\label{sec:int}

The gravitational interaction of orbiting objects with a background medium is 
fundamental in many astrophysical problems. Objects in gas-rich environments 
such as planets in protoplanetary disks, young stars in natal molecular clouds,
substellar mass objects in stellar winds, compact objects orbiting around 
supermassive black holes, and supermassive black hole binaries in colliding 
galaxies are subject to the gravitational interaction with background gas. 
The orbiting object naturally generates a gravitational density wake in its 
environment, inevitably followed by the reaction of the induced wake as 
dynamical friction. 

A recent series of studies on the gaseous dynamical friction is devoted 
to understand the orbital evolution of the object in rectilinear motion
\citep{ost99,san99,kim09,nam10,nam11} and in circular motion \citep{san01,
kim07,kim08,kwt10}. These theoretical and numerical works have covered the 
effects of binarity, nonlinearity, acceleration, and background stratification.
Albeit all these factors give rise to appreciable effects, the density 
wake and the dynamical friction force are primarily characterized by the 
accretional and orbital properties of the object described by the linear 
perturbation analyses in \citet{ost99} and \citet[hereafter \kk]{kim07}.

Given the importance of the gravitational drag, these linear analyses 
have been applied to various astrophysical aspects such as the orbital 
decay of planets around giant stars \citep{sch08,vil09}, the primordial 
mass segregation in massive star forming regions \citep{gar10,cha10}, the 
binary hardening in parent gaseous disk \citep{bar11}, the deceleration 
of galaxies as the heating sources of intracluster medium against the 
cooling flows \citep{elz04,kwt05,kwt07,con08}, and the coalescence of 
supermassive black holes in gas-rich mergers of galaxies \citep{esc04,
esc05,dot06,dot07,may07}. These linear analyses have been extended to 
the relativistic case \citep{bar07} for extreme mass-ratio inspirals in 
the vicinity of supermassive black holes. In particular, \citet{bar08} 
found an important difference in the inclination angle of the orbits 
between the dynamical friction and the radiation reaction possibly 
detectable by Laser Interferometer Space Antenna (LISA). 

Also important are the morphological and kinematic features of the density 
wake per se, as they can be directly observed by means of X-rays in shocked 
regions, molecular line emissions, dust continuum emissions, and in dust 
scattered light. \citet{fur02} raised the possibility that the nature of 
dark matter can be examined by detailed X-ray observations of wakes due to 
the motion of galaxies in the intracluster medium, exhibiting distinguishable 
pictures between the standard collisionless cold dark matter and the 
collisional (fluid) dark matter. \citet{san09} furthermore proposed that 
the morphology of a galactic wake in X-ray emission differs in purely 
baryonic Modified Newtonian Dynamics (MOND) from the case of Newtonian 
gravity with dark matter. These authors only inspected the local attributes of 
the gravitational density wakes using \citeauthor{ost99}'s \citeyearpar{ost99} 
formulae for the cases of the perturbing object in rectilinear motion.

However, many observable objects have fairly short orbital periods 
compared to the dispersion timescale of the background gaseous matter, 
leading to global shaping of the induced wakes. \kk\ \oops{showed that} 
the curvilinear orbit modifie\oops{s} not only the curvature of the 
wake but transforme\oops{s} the overall structure
entirely because of the memory effect\footnote{\ Memory effect (or battery
effect) originally refers, only in nickel cadmium rechargeable batteries, to 
the symptom of losing their maximum energy capacity after being repeatedly 
recharged to partially discharged states. We apply this term here for the wake
repetitively perturbed by the revolving object over the former perturbations 
remained, resulting in a loss of density contrast, as we will mention at the 
end of this paper.}. One application of the large scale wakes is to a case that
planets or brown dwarfs lie in cool giant envelopes, whose wakes are possibly 
observable at large distances from their bright parent stars. Future high 
sensitivity/resolution observations of these wakes can provide evidence of 
the unseen substellar mass objects, further allowing us to derive their 
orbital properties (\citealp{kim11} in prep.; for a brief introduction 
see \citealp{kim10}). 

Although \kk\ calculated the density wake, they 
primarily focused on the dynamical friction force that affects the 
initially circular orbit \oops{and the description of the general 
wake feature was somewhat superficial. A more detailed understanding 
of the wake characteristics is necessary in order to provide an 
interpretative framework for both further theoretical works under 
more complicated situations and future observations to detect the 
gravitational wakes of hidden objects}. In this paper, we extend 
the semianalytic work \oops{of \kk}\ highlighting the observable 
quantities, and determine the detailed structure and kinematics 
of the gravitational wake due to an object in circular motion. 
\oops{Particularly, we study the density structure in great detail 
based on an analytic linear analysis, which also helps to understand 
aspects of the simulated density wake in a nonlinear regime. We 
provide the density contrast of the wake as a function of the object 
mass, orbital radius, orbital speed, and the background sound speed, 
providing a way to obtain the information of unseen orbiting object
from the observation of the wake at a large distance.}

This paper is organized as follows. In \S\ref{sec:rev} we review the 
derivation and findings in previous works as the starting point of this 
investigation. In \S\ref{sec:lin} we present the nature of linear wake 
as revealed in the \oops{shape} (\S\ref{sec:mor}), the density enhancement 
(\S\ref{sec:den}), and the velocity distribution (\S\ref{sec:vel}), 
provided that the weak gravitational influence is thoroughly described 
in the linear perturbation analysis. In \S\ref{sec:non}, based on the 
linear wake properties, we suggest a new explanation for nonlinear 
effects manifested on the wake when the perturbation in discontinuity 
is saturated. In \S\ref{sec:sum} we summarize the newly found properties 
of the gravitationally induced wake of a circularly orbiting object in 
a static uniform gaseous medium, and briefly consider applications to 
a few astronomical objects.

\section{BACKGROUND REVIEW}\label{sec:rev}

Starting 
from the basic equations of hydrodynamics with an external time-dependent
gravitational potential $\Phi_p$ but excluding self-gravity of the gaseous 
medium, one can obtain a linearized three dimensional wave equation for the 
density enhancement $\alpha=\delta\rho/\rho$ under the assumption of a small 
perturbation ($\alpha\ll1$),
\begin{equation}\label{equ:wav}
  \vct{\nabla}^2\alpha-\frac{1}{\cs^2}
  \frac{\partial^2\alpha}{\partial t^2} = -\frac{1}{\cs^2}\nabla^2\Phi_p.
\end{equation}
Here, $\cs=(\gamma p/\rho)^{1/2}$ is the sound speed of unperturbed gas with
density $\rho$, thermal pressure $p$, and adiabatic index $\gamma$ in general. 
\citet{ost99} applied it to the situation that the perturbing object is in 
rectilinear motion through an infinite homogeneous gaseous medium, 
utilizing the retarded Green's function technique \citep{jac75} based on 
the Fourier transform \citep[see also][]{jus90}. A purely analytic solution 
for the density enhancement was derived, which can be expressed as
\begin{equation}\label{equ:ost}
  \alpha=\frac{r_B}{\left(d_\|^2+d_\bot^2(1-\mach^2)\right)^{1/2}}\ 
  \sum_{d_\|^{\,\prime}}\mathcal{H}\left(V_pt-d_\|^{\,\prime}\right)
\end{equation}
in the axisymmetric coordinates ($d_\|$, $d_\bot$) with the origin at the 
perturbing object, $d_\|$ along the tail (i.e., in the opposite direction of 
the orbital motion) and $d_\bot$ in the orthogonal direction. Here, we define 
the most important parameters in this paper, $r_B$ and $\mach$, as the Bondi 
accretion radius and the orbital Mach number of the object, respectively. 
In the summation of the Heaviside step functions $\mathcal{H}$, the primed
distance $d_\|^{\,\prime}$ denotes the original position of the gravitational 
perturbation, initiated by the object and currently arriving at the observing 
point ($d_\|$, $d_\bot$). 
Thus, the integrand means \oops{that} the perturbation sources should be 
on the trace of the object, inevitably not able to be further than the 
initial position of the object, $V_pt\ge d_\|^{\,\prime}$. The summation 
of only the step functions indicates the exactly same contributions from 
the individual perturbations gathered at a time. Accordingly, it can be 
substituted simply by the number of perturbations at the point,
$\pert=\sum_{d_\|^{\,\prime}}\mathcal{H}\,(V_pt-d_\|^{\,\prime})$, 
which divides the space into three regions showing discontinuous density 
and velocity distributions: $\pert=2$, 1, and 0 in the cone region, in 
the ice cream region, and outside of these two regions (as following the 
author's original nomenclature). These three regions are separated by 
the sharp edge of a Mach cone and the modest edge of a sonic sphere about 
the initial position of the object. As the orbital Mach number $\mach$ 
decreases, the apex of the Mach cone attached to the object approaches to, 
and finally disappears in, the sonic sphere with the transition occurring 
at the sonic speed \oops{(}$\mach=1$\oops{)}, eliminating the shock front 
from the density wake.

Using the same retarded Green's function technique, \kk\ applied the three 
dimensional wave equation (eq.~[\ref{equ:wav}]) to the linear wake induced 
by a circularly orbiting object at distance $r_p$ from the system center 
in a static uniform gaseous background. The resulting density enhancement 
is given as
\begin{equation}\label{equ:k07}
  \alpha=\frac{r_B}{r_p}\sum_{\varphi^\prime}
  \frac{\mathcal{H}\left(V_pt/r_p-\varphi^\prime\right)}{\left|
    \mach^{-1}\varphi^\prime-\mach(r/r_p)\sin(\varphi^\prime-\varphi)\right|}
\end{equation}
in cylindrical coordinates\footnote{\ $\varphi$ is the angular distance from 
the object in the opposite direction of the orbital motion, tracing the induced
wake tail, i.e., the perturbing object has moved from $V_pt/r_p$\,($>$0) to 0. 
This is corresponding to $-s$ in the notation of \kk.} $(r,\,\varphi,\,z)$ with
the equatorial plane aligned to the orbital plane of the perturbing object. All
the possible perturbations must be launched from the object passing through the
locations $(r_p,\,\varphi^\prime,\,0)$, where $\varphi^\prime$ satisfies the 
condition,
\begin{equation}\label{equ:sol}
  \mach^{-1}\varphi^\prime=\Big(1+(r/r_p)^2+(z/r_p)^2
  -2(r/r_p)\cos(\varphi^\prime-\varphi)\Big)^{1/2}.
\end{equation}
This relation describes that the individual perturbation propagates from 
the launching point $(r_p,\,\varphi^\prime,\,0)$ to the observing point 
$(r,\,\varphi,\,z)$ as a sound wave during the time that the perturbing 
object travels the angular distance $r_p\varphi^\prime$ along the orbit with 
the speed $\mach$. In the case of curvilinear motion, the summation is not 
only restricted to the step function, implying different contributions from 
the different elements of the perturbations. This diversity of contributions 
is the primary reason of the complications arising in a density wake of the 
circularly orbiting object.

\kk\ accurately calculated the wake density in equation (\ref{equ:k07}) by 
computing the positions of perturbation origins in the nonlinear equation 
(\ref{equ:sol}) using root-finding algorithms based on either Newton-Raphson 
or bisection iterations. Their accurate numerical values, however, do not 
provide deep \oops{physical} insight for the global features of the induced 
density wake, 
which are useful to probe the orbital properties of the perturbing object. 
In this paper, we revisit \kk's work on the linear density wake of 
a circularly orbiting object by modifying their analytic formulae and by 
inspecting the global distributions of the calculated density enhancement 
and velocity fields, if necessary, in comparison with the linear trajectory 
counterparts. In addition, we have performed high resolution three dimensional 
hydrodynamical simulations using an adaptive mesh refinement code, FLASH 
\citep{fry00}, to investigate the response of a gaseous medium to the 
influence of an extended object, characterized by a gravitational 
softening radius $r_s$ in the range of 0.01--0.1\,$r_p$ (alternatively, 
0.1--25\,$r_B$) for $\mach=0.2$--10. This parameter space includes the 
nonlinear regime showing different behaviors of the wake from the ones 
in the linear analysis \citep[see also][]{kim09,kwt10}. All the simulations 
treat only a pseudo-isothermal gas with $\gamma=1.00001$ in the FLASH code, 
providing a constant sound speed $\cs$ in the entire computational domain.

\section{LINEAR WAKE}\label{sec:lin}

\subsection{Morphology}\label{sec:mor}

An object orbiting with a subsonic speed ($\mach<1$) creates a smoothly 
distributed wake without a shock, showing a gradual decrease of density 
along the distance from the perturbing object. The isodensity surfaces of 
this subsonic wake have the shapes of oblate spheroids with ellipticity 
$e=\mach$ near the perturbing object (see Appendix A of \kk\ for derivation), 
gradually twisted in a trailing fashion with the low-density outermost 
shells tending toward comma shapes. The morphology of a subsonic wake is 
simple and easily understood since the wake receives only one sonic 
perturbation everywhere, contrary to the supersonic wakes ($\mach>1$) 
investigated in depth in this section.

The wake induced by a circularly orbiting object with supersonic speed 
has \oops{a spiral-}onion shell structure in three dimensional space, 
having distinct boundaries for the high density region and the lower 
density inter-structure regions. The density contrast is primarily 
caused by the piling up of the delayed gravitational perturbations inside 
the high density structure. Equation (\ref{equ:sol}) gives the positions 
of the origins, $\varphi^{\,\prime}$, for the perturbations reaching 
the observing point $(r,\,\varphi,\,z)$ at time $t$. After some algebra 
from equation (\ref{equ:sol}), \kk\ showed \oops{that} the high density 
structure possesses $\pert=2n+1$ perturbations, where $n$ is an integer number 
depending on the orbital Mach number such as 1 ($\pert=3$) at $\mach\sim1-4.6$,
2 ($\pert=5$) at $\mach\sim4.6-7.8$, 3 ($\pert=7$) at $\mach\sim7.8-10.9$, and 
so on (see Appendix B of \kk\ for a more detailed discussion). For comparison, 
in subsonic cases ($\mach<1$) the wake receives one perturbation ($\pert=1$) 
everywhere, corresponding to $n=0$.

Extending the analysis in Appendix B of \kk, in which the number of 
perturbations \oops{of} the arm structure in the orbital plane is obtained, 
we derive a purely analytic formula to describe the exact shape of the 
enhanced density structure in three dimensional space. The details of 
the derivation are omitted but the essence is to use the property at 
the boundaries, where the number of perturbations changes so that 
the curve on the right hand side of equation (\ref{equ:sol}) has the 
same slope as the line on the left hand side. \oops{W}e derive an 
expression for the shape of the enhanced density structure given by
\begin{equation}\label{equ:mor}
  \cos\left(\left|(X^2-1)^{1/2}\mp(Y^2-1)^{1/2}\right|-\varphi\right)
  =\frac{1\pm(X^2-1)^{1/2}(Y^2-1)^{1/2}}{XY}
\end{equation}
for the outer ({\it upper sign}) and inner ({\it lower sign}) boundaries, 
where $X$ and $Y$ are defined as $X=\mach(d_++d_-)/2$ and $Y=\mach(d_+-d_-)/2$.
Here, the normalized distances $d_-=\left((r-r_p)^2+z^2\right)^{1/2}/r_p$ from 
the perturbing object and $d_+=\left((r+r_p)^2+z^2\right)^{1/2}/r_p$ from the 
mirror point about the system center. The solutions of these formulae determine
the boundaries of the high density region, revealed in Figure~\ref{fig:ptm}a--b
as a single-armed spiral in the orbital plane ($z=0$) and arcs in a meridional 
plane ($\varphi=0$). 

\subsubsection{Spiral}\label{sec:spi}

\oops{Shaping a spiral arm having a constant width is physically not 
so trivial for the density wake in a uniform background. Because of 
the gas pressure, the gaseous matter accumulated behind the object 
tends to disperse with the sound speed in a static uniform background.
Thus, the excess density region should form an extending tail similarly
to the Mach cone shape in the linear trajectory counterpart, unless 
the background density configuration is changed. As we will show in 
\S\ref{sec:bac}, the initially uniform background transforms the
distribution into a hydrodynamic equilibrium about the system center.
As the result of the centrally concentrated background, the inner part
of the spiral does not propagate inward but outward, following its outer
part.}

\oops{In this section, we construct an explicit form of the spiral shape,
which is important in order to derive the pitch angle, the arm width,
and the spacing of the spiral pattern, relevant to the object orbital
parameters.} Equation (\ref{equ:mor}) at $z=0$ \oops{yields} the polar
equations \oops{for} the boundaries \oops{of} the spiral arm in the 
orbital plane (see also eq.~[B2] in \kk),
\begin{eqnarray}\label{equ:spi}
  \varphi=\left|\Big(\mach^2(r/r_p)^2-1\Big)^{1/2}\mp
  \left(\mach^2-1\right)^{1/2}\right|-\cos^{-1}\left(
  \frac{1\pm\left(\mach^2-1\right)^{1/2}\Big(\mach^2(r/r_p)^2-1\Big)^{1/2}}
	  {\mach^2(r/r_p)}\right) \quad {\rm at}\ \ z=0,
\end{eqnarray}
in which the upper and lower signs are for the outer and inner boundaries, 
respectively. The outer boundary of the arm, originated from the Mach cone 
in \citet{ost99}, lies not only outside the orbit ($r/r_p\geq1$) but it also 
extends to $r/r_p=\mach^{-1}$ (see Fig.~1 in \kk). At this innermost point of 
the outer boundary, the sonic sphere due to the influence of the gravitational 
perturbation from the object at the initial location starts to overlap with 
the curved Mach cone tailing the object, forming the inner boundary of the 
spiral arm.

In order to see the 
global features of the wake in the orbital plane, we differentiate the 
complicated forms (eq.\,[\ref{equ:spi}]) following the exact shapes. 
Using the angle transformation formulae for cosine, it is reduced to
\begin{equation}\label{equ:tan}
  \frac{d(r/r_p)}{d\varphi}=\frac{r/r_p}{\Big(\mach^2 (r/r_p)^2-1\Big)^{1/2}}
\end{equation}
for both outer and inner arm boundaries in the orbital plane, except for 
a small part of the outer boundary inside the orbit ($\mach^{-1}<r/r_p<1$)
having the opposite sign. We define an angle $\Theta$ as its tangent value,
$\tan\Theta$, corresponding to equation (\ref{equ:tan}), which leads to the
alternative form
\begin{equation}\label{equ:sin}
  \sin\Theta=\frac{r/r_p}{\Big((\mach^2+1)(r/r_p)^2-1\Big)^{1/2}},
\end{equation}
resulting in the familiar result $\Theta=\sin^{-1}\mach^{-1}$ at the object 
position $r/r_p=1$ as the half-opening angle of the curved Mach cone (\kk; 
see also \citealp{ost99}). This angle $\Theta$ decreases along the distance 
$r/r_p$ as seen in equations (\ref{equ:tan}) and (\ref{equ:sin}), but it 
converges very quickly to $\Theta=\tan^{-1}\mach^{-1}$, showing a small 
deviation of $\le$~1\degree\ at $r/r_p\ge2$ for $\mach\ge2$ ($r/r_p\ge4$ 
for $\mach=1$). This quasi-constancy of $\Theta$ suggests that the density 
wake in a static gaseous medium indeed approximates an Archimedes spiral, 
satisfying $r/r_p=\varphi\mach^{-1}+C$ where the constant $C$ differs 
between the inner and outer boundaries of the high density arm region. 
Figure~\ref{fig:ptm}a displays the density enhancement in the orbital plane 
($z=0$) of the perturbing object, and exhibits a high density region enclosed 
by two Archimedes spirals shown in dashed ($C=1$) and dotted ($C=-0.7$ for 
$\mach=10$) lines. The constant $C$'s are calculated at $r/r_p=1$ using 
equation (\ref{equ:spi}); the value for the inner boundary changes with 
$\mach$, while it is unity for the outer boundary.

\subsubsection{Arcs}\label{sec:arc}

An investigation of the wake shape in \oops{a} meridional plane is needed to
provide an interpretative framework for edge-on observations. \oops{Compared 
to the asymmetric spiral pattern seen in the face-on line of sight, this arc
pattern is not properly spotlighted and hence not analyzed well. In this
section, we show the purely analytic solution for the arc shape of the
gravitational wake, and use a geometrical approach to fully understand
the wake shape relevant to the object properties.}

A trivial solution of equation (\ref{equ:mor}) is both $X=1$ and $Y=1$ 
in the meridional plane wherein the perturbing object lies ($\varphi=0$), 
which indicates the (vertically) highest intersecting points between the 
outer and inner boundaries, i.e., the extension limit of the intensively 
perturbed region. $X=1$ and $Y=1$ stand for an ellipse and a hyperbola, 
by definition, with the common foci at the location of the perturber and 
the mirror point about the orbital axis. However, both $X=1$ and $Y=1$ 
actually correspond to the same formula for a hyperbola
\begin{equation}\label{equ:hyp}
  \mach^2(r/r_p)^2-\frac{\mach^2}{\mach^2-1}(z/r_p)^2=1,
\end{equation}
since we only consider the shape of a wake induced by a supersonic perturber 
($\mach>1$). The vertex of the hyperbola, $r/r_p=\mach^{-1}$, corresponds to 
the distance of the junction between the outer and inner arm boundaries in the 
orbital plane. The extension limit of the high density arcs in the vertical 
direction is simplified to the asymptotes of the hyperbola, 
\begin{equation}\label{equ:asy}
  z=\pm\,(\mach^2-1)^{1/2}\,r,
\end{equation}
which are overlaid by black lines in Figure~\ref{fig:ptm}b. This indicates 
that the angular size of the arcs is $2\tan^{-1}(\mach^2-1)^{1/2}$, further 
extended with higher orbital Mach number. This vertical stretch, however, does 
not depend very strongly on $\mach$ except for near unity (i.e., 60\degree-- 
90\degree\ for $\mach\ge2$).

In Figure~\ref{fig:ptm}b individual segments of the boundaries, 
satisfying equation (\ref{equ:mor}), appear to be circular arcs 
about the centers at either the object position or the mirror point. 
In the plane of the object ($\varphi=0$; $x>0$), the arcs extended 
from the outer boundaries of the spiral arm in Figure~\ref{fig:ptm}a 
(i.e., $x/r_p=1$, $1+2\pi\mach^{-1}$, $1+4\pi\mach^{-1}$, 
$1+6\pi\mach^{-1}$, $1+8\pi\mach^{-1}$, and $1+10\pi\mach^{-1}$ at 
$z/r_p=0$) center around the perturbing object at $(x,z)/r_p=(1,0)$. 
These concentric arcs have endpoints on the hyperbola defined in equation 
(\ref{equ:hyp}) (or, the asymptotes in eq.~[\ref{equ:asy}]), except for
the point (not visible) at the perturber position and the curve passing 
through $(x,z)/r_p=(1+2\pi\mach^{-1},0)$, which completes its full circle. 
The inner $x$-intercept of this circle at $x/r_p=1-2\pi\mach^{-1}$ actually 
corresponds to the inner edge of the curved Mach cone, which is considered 
as a part of the outer arm boundary (see \S\ref{sec:spi}) and, here, this 
is clarified. On the other hand, the arcs crossing the inner arm boundaries 
on this side ($x>0$) have the center at the mirror point $(x,z)/r_p=(-1,0)$, 
resulting in larger radii compared to the radii of the outer counterparts. The 
outer boundaries on the opposite side ($x<0$), i.e., the arcs having intercepts
at $x/r_p=-1-\pi\mach^{-1}$, $-1-3\pi\mach^{-1}$, $-1-5\pi\mach^{-1}$, 
$-1-7\pi\mach^{-1}$, and $-1-9\pi\mach^{-1}$, are concentric about the 
mirror point of the object, $(x,z)/r_p=(-1,0)$. On this side ($x<0$),
instead, the center of the inner boundaries is at the object position,
$(x,z)/r_p=(1,0)$.

The inner and outer boundaries, in fact, constitute the same circles, which 
is better visualized in Figure~\ref{fig:num} for the number of perturbations.
Particularly, the largest sphere about $(x,y,z)/r_p=(1,0,0)$ limits the region
already in a steady state, receiving 1, 3, 5, and 7 perturbations in different 
localities. The arcs (or circles) in Figure~\ref{fig:num}b are formalized by 
equation (\ref{equ:sol})\footnote{\ Eq.~(\ref{equ:mor}) gives the same results.
Since $|(X^2-1)^{1/2}\mp(Y^2-1)^{1/2}|$ in this equation is originated from 
$\varphi^\prime$ in eq.~(\ref{equ:sol}), eq.~(\ref{equ:mor}) is equivalent
to $(X^2+Y^2-2XY\cos(\varphi^\prime-\varphi))^{1/2}=\varphi^\prime$. In the 
meridional plane of $\varphi=0$, the gravitational perturbations from the 
object located, in the past, at $\varphi^\prime=2\pi\times$0 ({\it current}), 
1, 2, 3, 4, and 5 ({\it initial}) form the arcs as the segments of circles,
$d_-=(X-Y)\mach^{-1}=2\pi\mach^{-1}\times$0, 1, 2, 3, 4, and 5. Similarly,
$\varphi^\prime=\pi\times$1, 3, 5, 7, and 9 correspond to the concentric 
arcs on $d_+=(X+Y)\mach^{-1}=\pi\mach^{-1}\times$1, 3, 5, 7, and 9.} with 
the sources at $\varphi^\prime$ of $\pi$-multiples in the plane of $\varphi=0$.
The perturbation launched every time the object returns back to the original 
position ($\varphi^\prime$ being a multiple of $2\pi$) propagates as a sound 
wave leaving the vestige on the spherical shell with the radius of multiple of
$2\pi\mach^{-1}$; the perturbation generated half an orbit ($\pi$ in angular
phase) later is marked on the sphere, smaller by $\pi\mach^{-1}$, with the
center at the mirror point in the same meridional plane. And by extension, the 
spiral in the orbital plane (Fig.~\ref{fig:num}a) emerges as an aftermath of 
the phase difference of the perturbation source.

\subsection{Density Enhancement}\label{sec:den}

The density enhancement, $\alpha$, in the pattern follows a complicated 
rule because of the accumulation of gravitational perturbations launched 
from the object at different positions on the curvilinear orbit. This 
complication is particularly caused by the different contributions of 
the respective perturbations, unlike the case when the object is in a 
straight-line trajectory. The density enhancement profiles are displayed 
in Figure~\ref{fig:ptm}c as a function of radial distance in various 
directions along the spiral arm (the lines in red, yellow, green, and 
blue colors) and along the orbital axis (the cyan line), in which all 
the curves appear well located within the black lines (see below). The 
profiles in the orbital plane reveal the highest density peaks at the 
object orbit, $r/r_p=1$, and the peak values decrease with distance from 
the orbit. The minimum density enhancement in azimuthal directions, i.e.,
the baseline of the group of profiles, gradually decreases following the 
bottom dashed line (up to $r/r_p=0.4$) starting from $\alpha=r_B/r_p$ at 
the orbital center. At $r/r_p=0.4$, however, the baseline jumps to, and 
follows, another dashed line, and another jump occurs at $r/r_p=0.75$. 
In this example case of $\mach=10$, the baseline moves its paths among 
the dashed lines through three steps before reaching the orbital radius. 
Beyond the orbital radius, it follows the third dashed line from bottom 
until it abruptly drops at $r/r_p\sim2.5$ apart from the path. This last 
drop of minimum density enhancement coincides with the sudden disappearance 
of the inner arm boundary, which is related to the orbital evolution time 
of the object. Using the spiral functional forms deduced in \S\ref{sec:spi}, 
it is shown that the outer and inner arm boundaries reach $r/r_p=4.1$ and 
2.4, respectively, at the moment that the perturber completes 5 orbits 
(i.e., $\varphi=10\pi$). The inner arm boundary lags behind by 2.7-turn 
since it starts from $(r/r_p,\,\varphi)=(\mach^{-1},\,1+0.7\mach)$, 
deeper inside than the starting point of the outer boundary at 
$(r/r_p,\,\varphi)=(1,\,0)$.

\subsubsection{Interarm Region}\label{sec:bac}

\begin{figure}
  \epsscale{0.95}
  \plotone{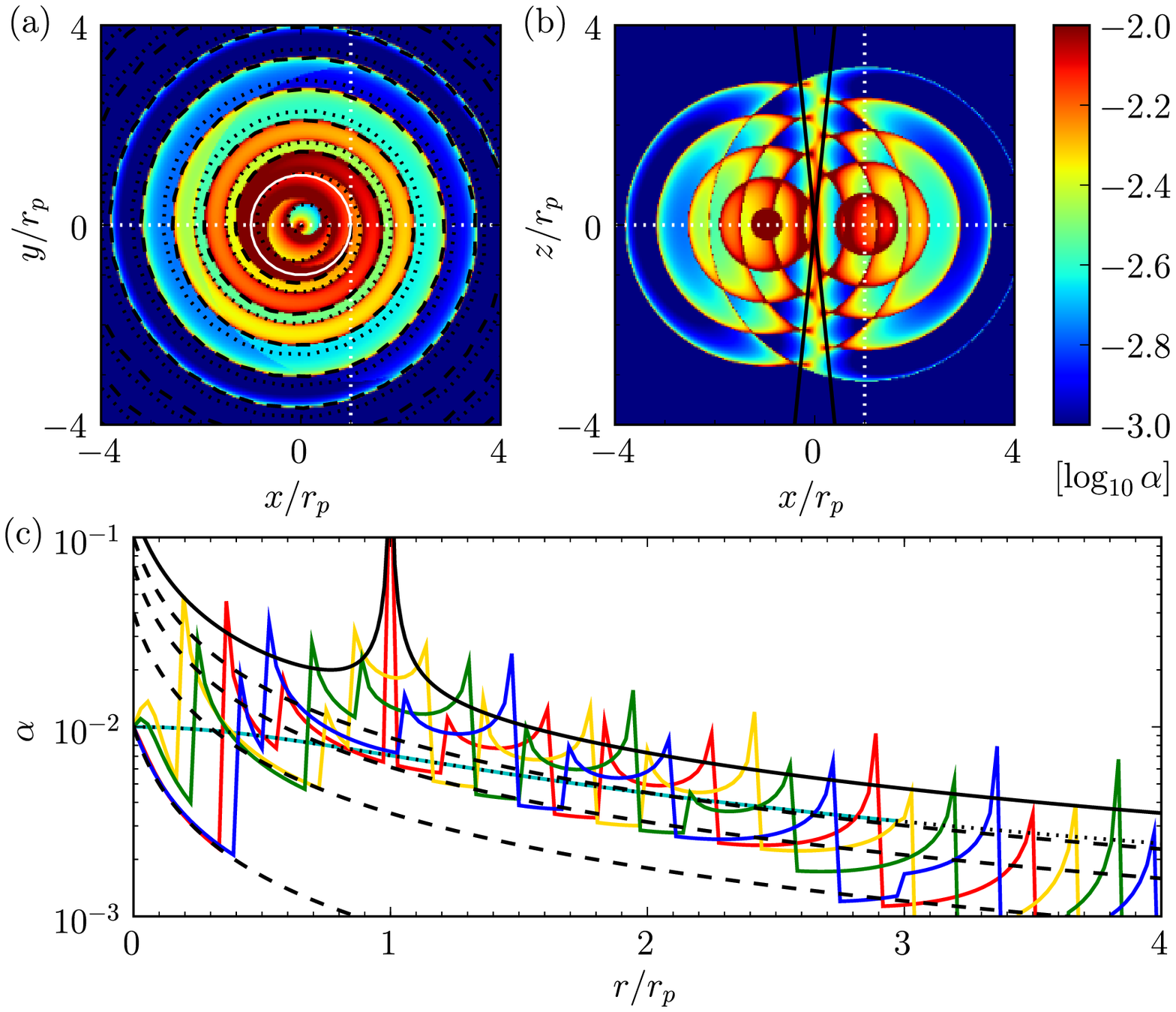}
  \caption{\label{fig:ptm}
    Density enhancement $\alpha$ of the spiral-onion shell shaped density wake 
    generated by a point mass object orbiting at a fixed distance $r_p$ with 
    an orbital Mach number $\mach=10$ in an initially static uniform gaseous 
    medium after 5 orbital periods, calculated by a semianalytic method in \kk.
    (a) In the orbital plane, the induced density wake possesses a condensed 
    trailing arm bounded by outer ({\it black dashed}) and inner ({\it black
    dotted}) Archimedes spirals $r/r_p=\varphi/\mach+C$ with $C=1$ and $-0.7$, 
    respectively, in the polar coordinates ($r$, $\varphi$) with $\varphi$ in 
    the clockwise direction tracing the spiral but opposed from the motion of 
    the object. White circle represents the circular orbit of the object, 
    currently located at the intersection of white dotted lines. (b) In a 
    meridional plane, the high density area of the wake displays arcs being 
    able to extend in angle of $\tan^{-1}(\mach^2-1)^{1/2}$ from the orbital 
    plane ({\it black solid}). White dotted line indicates the current location
    of the object as in (a). (c) The density enhancement profiles as a function
    of the distance from the center are displayed along ${}^{+}x$ ({\it red}), 
    ${}^{-}y$ ({\it yellow}), ${}^{-}x$ ({\it green}), ${}^{+}y$ ({\it blue}), 
    and $z$ ({\it cyan}) axes. Notice that $\alpha=r_B/r_p$ at the center, 
    where $r_B$ is the Bondi accretion radius of the object. The black solid 
    line ($\alpha=\alpha_1+14\amin$) outlines the peaks at the spiral arm 
    boundaries, the dashed lines ($\alpha=\amin\times1,4,7,10$ from bottom) 
    show the change of minimum density enhancement in the orbital plane 
    because of piling-up of delayed perturbations, and the dotted line
    represents well the density enhancement of $\amin$ along the orbital axis. 
    See text for more comprehensive explanation including the definition of 
    $\alpha_1$ and $\amin$.
  }
\end{figure}

\begin{figure}
  \plotone{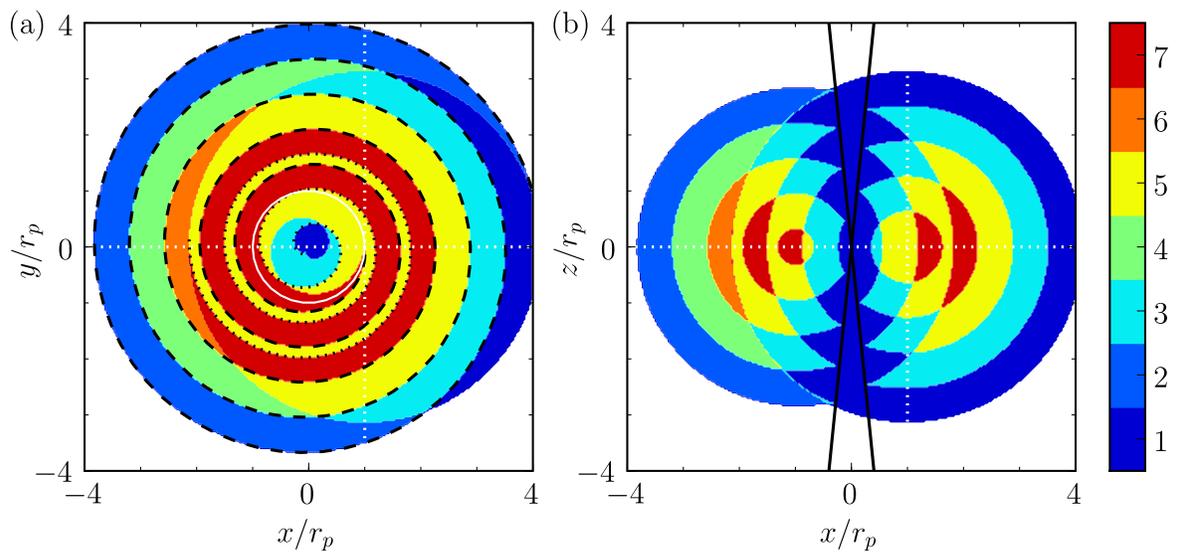}
  \caption{\label{fig:num}
    Same as Fig.~\ref{fig:ptm}a--b, but for the number of perturbations. 
    The perturbing object has circled 5 times with an orbital Mach number 
    of $\mach=10$. 
  }
\end{figure}

Equations (\ref{equ:k07})--(\ref{equ:sol}) indicate that the minimum of density
enhancement in the azimuthal direction satisfies\footnote{\ Strictly,
eqs.~(\ref{equ:k07})--(\ref{equ:sol}) imply the minimum value of density 
enhancement in the central part $r/r_p<2\mach^{-2}$ is better represented by 
$\alpha$ $\simeq$ $r_B/\left((r+r_p)^2+z^2\right)^{1/2}$ than $\amin$ defined 
in eq.~(\ref{equ:alm}). This is simply seen in an extreme case of $\mach=0$ 
so that $2\mach^{-2}$ goes to infinity; its density wake in hydrostatic 
equilibrium about the static perturber at $(r,\,\varphi,\,z)=(r_p,0,0)$ 
shows the density enhancement $\alpha=\exp(r_B/d)-1\approx r_B/d$ (when 
$r_B/d\ll2$) at the distance $d=(r^2+z^2+r_p^2-2r r_p\cos\varphi)^{1/2}$ 
from the perturber, which has the minimum value at $\varphi=\pi$.}
\begin{equation}\label{equ:alm}
  \amin=\frac{r_B}{(r^2+z^2+r_p^2)^{1/2}+r\mach}.
\end{equation}
This is plotted in Figure~\ref{fig:ptm}c by the bottom dashed line, 
outlining the interarm density enhancement for all directions.
The incremental steps, observed above, follow the dashed lines 
exhibiting $\alpha=\amin(r,\,z=0)\times1,4,7,\cdots,1+3n$, where 
$n=(\pert-1)/2$ is 0, 1, 2, etc.\ depending on the orbital Mach number 
$\mach$ (for $n$ see \S\ref{sec:mor}, originated by eq.~[B3] in \kk). 

\begin{figure}
  \plotone{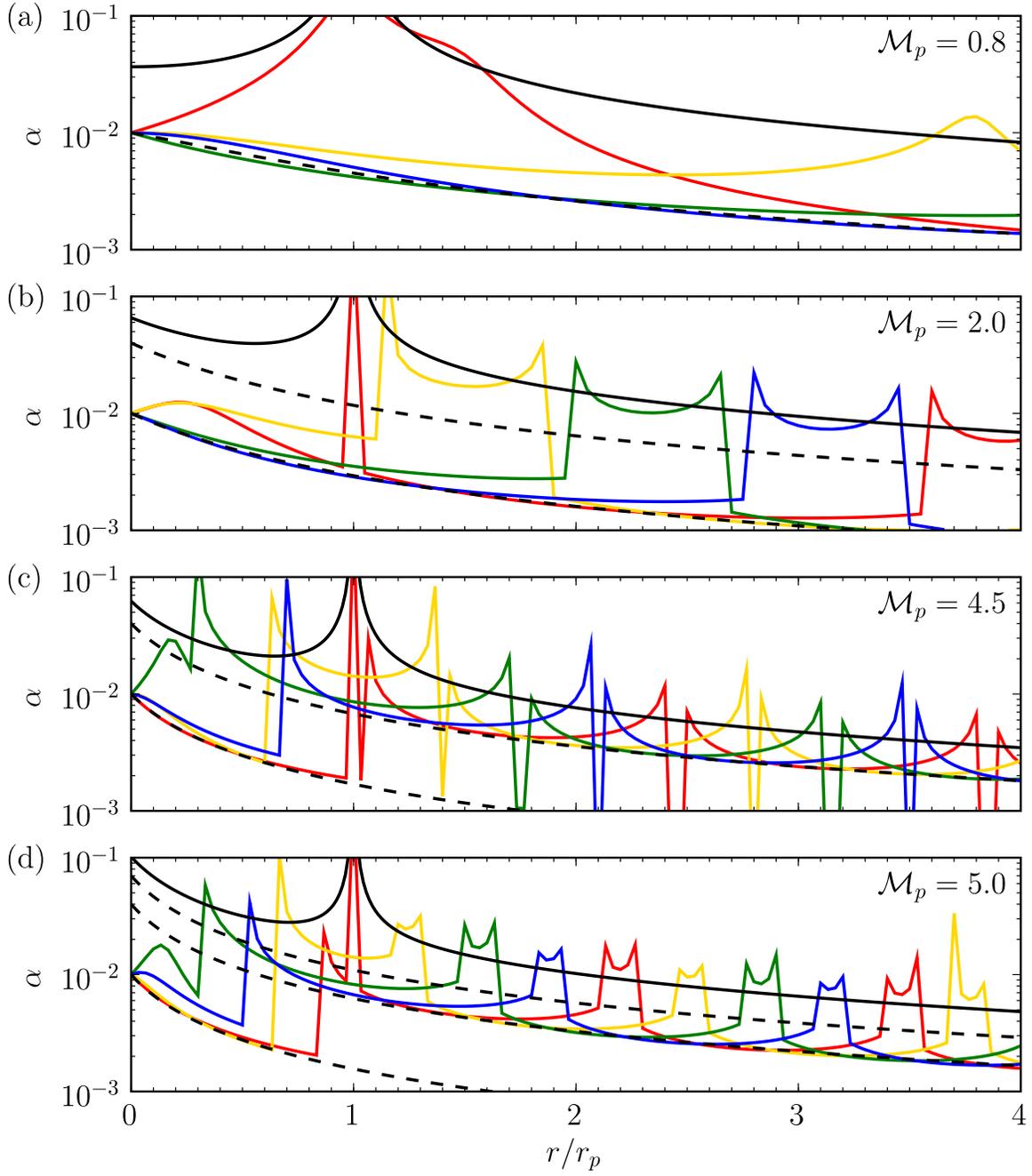}
  \caption{\label{fig:cmp}
    Same as Fig.~\ref{fig:ptm}c, but for (a) $\mach=0.8$ (b) $\mach=2.0$, 
    (c) $\mach=4.5$, and (d) $\mach=5.0$. In each panel, the $n+1$ dashed
    and one solid lines represent $\alpha=\amin\times1,4,\cdots,1+3n$ from 
    bottom and $\alpha=\alpha_1+(4n+2)\;\amin$, respectively, in the orbital
    plane with $n$ of (a) 0, (b) 1, (c) 1, and (d) 2. 
    See text for the dependence of $n$ on the orbital Mach number $\mach$,
    and the definitions of $\amin$ and $\alpha_1$.
  }
\end{figure}

To examine the dependence of the minimum density enhancement on $\mach$ 
(technically, $n$), we compare the $\alpha$-profiles for $\mach$ of 0.8 
($n=0$), 2.0 ($n=1$), 4.5 ($n=1$), and 5.0 ($n=2$) in Figure~\ref{fig:cmp},
having $n+1$ dashed lines in each panel. Unlike the subsonic case ($n=0$) 
having smooth $\alpha$-profiles above $\amin$ in Figure~\ref{fig:cmp}a, the 
supersonic cases ($n\geq1$) in Figure~\ref{fig:cmp}b-d show arm structures 
with sharp boundaries of higher density than in the middle of the arm. In 
Figure~\ref{fig:cmp}b for $\mach=2$ ($n=1$), the interarm density is still 
represented by $\amin$ ({\it lower dashed}) and the local minimum inside the 
arm is higher than $4\amin$ ({\it upper dashed}). Increasing $\mach$ (up to 
4.6) increases the arm width with a corresponding decrease in interarm extent 
(see Fig.~\ref{fig:cmp}c). The minimum arm densities concurrently approach 
$4\amin$. At the critical Mach number $\mach=4.6$, the outer boundary of 
the arm meets with the inner boundary of the arm one turn ahead, leaving no 
gap between arms. Because of this overlap (outside the orbit), the colored 
lines in Figure~\ref{fig:cmp}d for $\mach$ beyond 4.6 no longer exist on 
the bottom dashed line ($\alpha=\amin$). Instead, the minimum density is 
now given by the value of $\alpha=4\amin$, characterized as the local density 
minimum within the (infinitely) broad arm, above which a new narrow arm forms. 
The new arm possesses an increased number $\pert$ of perturbations, 
changing $n$ from 1 to 2, because this arm is basically the overlapping 
region. Increasing $\mach$ above 4.6 repeats the same broadening of the 
new arm until $\mach$ reaches another critical Mach number, 7.8. Above 
this critical Mach number, an overlapping event causes another jump of the 
minimum density to $(1+3(n-1))\amin$ with the increased $n$ ($\geq$1, in 
general), and leads to the density of the newly formed arm above $(1+3n)\amin$.

As shown in Figure~\ref{fig:ptm}b, the density structure at high latitudes 
($|z|>0$) is more complicated because of the partial overlaps of overdense 
areas closed by arcs with different radii. The overdense regions induced 
by a slower object ($1<\mach<4.6$) are composed of simple crescent-shapes 
without overlap. However, once overlapped ($\mach\geq4.6$), the regions of 
interest exhibit enhanced densities because they swallow the perturbations 
of each crescent-shaped structures. 
For instance, the case in Figure~\ref{fig:ptm} ($\mach=10$) experiences 3 
steps of density enhancement (corresponding to the number of perturbations
$\pert=7$) within the discernible spiral arm in the orbital plane. But the 
number of perturbations decreases at high latitude where the overlapping 
events are reduced (see also Fig.~\ref{fig:num}b). Eventually above the 
extension limit (Eq.~[\ref{equ:hyp}]), the gas flow experiences only one 
perturbation so that the density enhancement has the value of $\amin(r,z)$ 
(see cyan line in Figure~\ref{fig:ptm}c along the dotted line, 
$\alpha=\amin(r=0,\,z)$).

\subsubsection{Arm Region}\label{sec:arm}

In order to understand the distribution of the peak densities along distance, 
it is preferable to refer to the simpler case of the perturbing object moving 
on a linear trajectory, whose respective perturbations equally contribute to 
the density enhancement of the induced wake. \citeauthor{ost99}'s formula 
for the linear motion of a perturber (eq.~[\ref{equ:ost}]) has roughly an 
anticorrelation with respect to the distance from the perturbing object with 
the amplitude of $r_B\,|\mach^2-1|^{-1/2}$ per each contribution. Modifying 
\citeauthor{ost99}'s formula, for circular orbit cases we hypothesize a 
density enhancement due to one perturbation,
\begin{equation}\label{equ:al1}
  \alpha_1=\frac{r_B}{|r-r_p|}\frac{1}{|\mach^2-1|^{1/2}}
\end{equation}
in the orbital plane. However, in reality, the circular motion modifies the 
density enhancement from a simple collection of $\alpha_1$ by the number of 
perturbations. Particularly, gas pressure at the central region prevents the 
propagation of Mach waves, forming instead an equilibrium state with the 
central value $\alpha=r_B/r_p$. A rough guess of $\alpha_1$ can be used 
in an empirical formula for the peak density enhancement in the spiral arm 
pattern,
\begin{equation}\label{equ:alp}
  \alpha\approx\alpha_1+2\pert\amin,
\end{equation}
which shows fairly good fits in Figures~\ref{fig:ptm}c and \ref{fig:cmp} (see
black solid curve), except at the exact locations of the arm boundaries where
infinite values are introduced by the point mass perturber (see below).

\subsubsection{Effect of Object Size}\label{sec:siz}

\begin{figure}
  \plotone{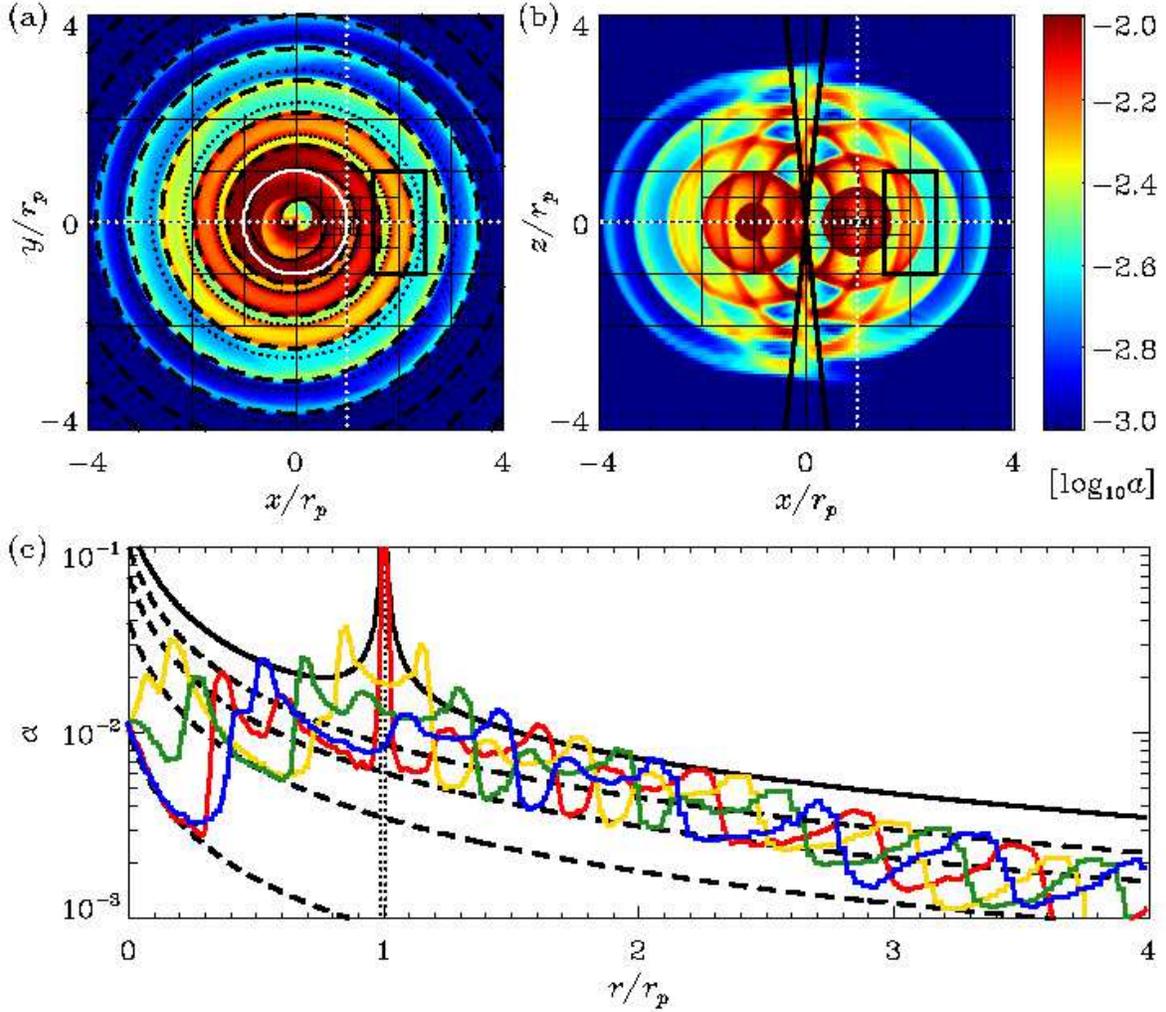}
  \caption{\label{fig:rss}
    Density enhancement $\alpha$ of the spiral-onion shell density wake 
    driven by an extended perturber with the gravitational softening radius
    of $r_s=0.01r_p$ in the unit of the orbital radius $r_p$, calculated by 
    a numerical simulation using an adaptive mesh refinement FLASH code with 
    the maximum refinement level 8 corresponding to the spatial resolution of 
    $0.001r_p$ (as indicated by thin black boxes including $64\times64\times32$
    meshes inside). See Fig.~\ref{fig:ptm} for the details. The thick black 
    boxes in (a) and (b) denote the regions plotted in Fig.~\ref{fig:vel}, and
    the vertical dotted lines in (c) represent the gravitational size $r_s$ of 
    the object.
  }
\end{figure}

\begin{figure}
  \plotone{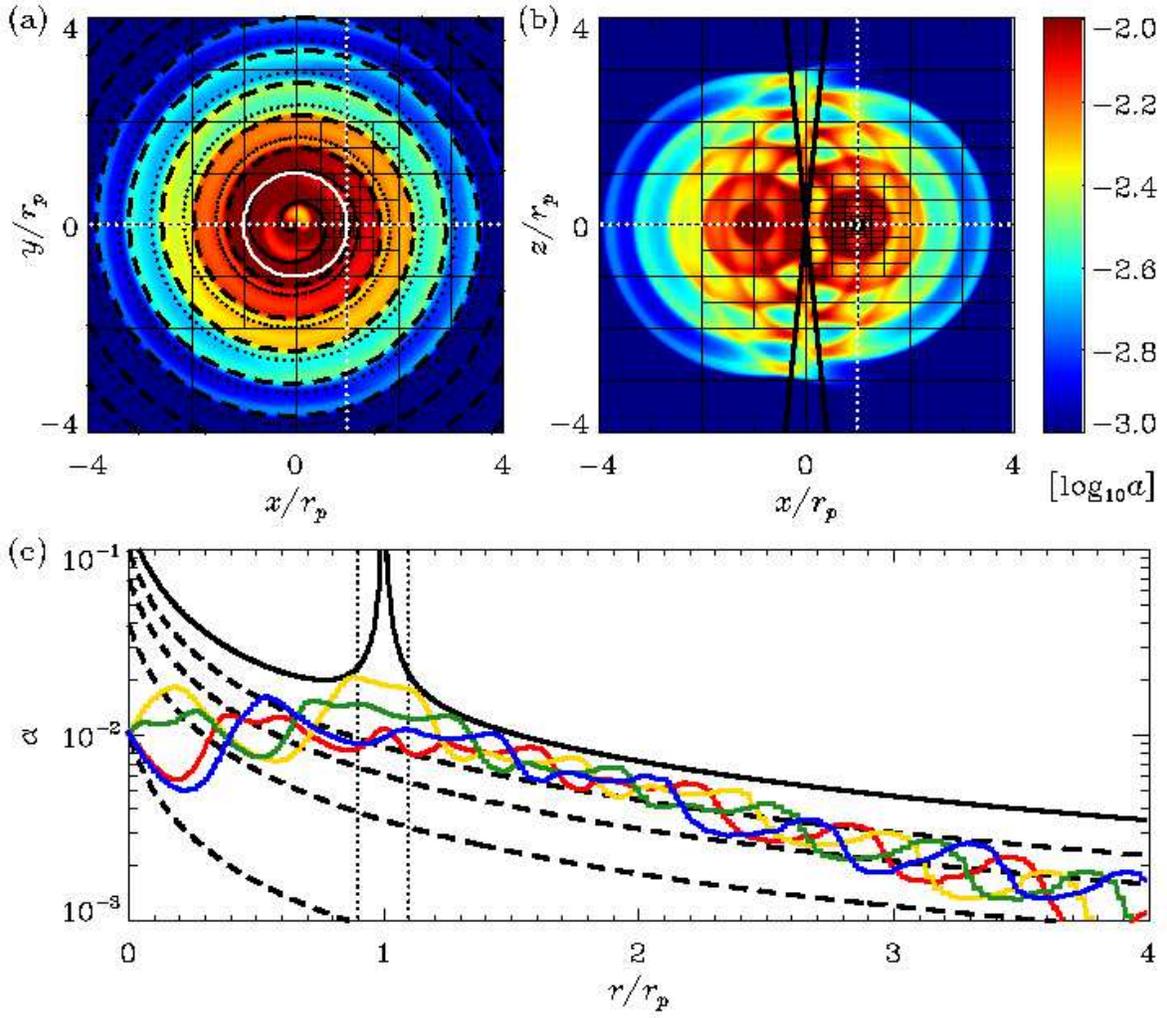}
  \caption{\label{fig:rsl}
    Same as Fig.~\ref{fig:rss}, but for a more extended perturber with the 
    gravitational softening radius of $r_s=0.1r_p$, comparable to half 
    ($0.09r_p$) of the interarm spacing in the case of a point mass perturber.
  }
\end{figure}

Although, mathematically, the perturbation from a point mass causes 
infinite density at the exact locations of arm boundaries, it is 
limited to the immediate vicinity of the arm boundaries, detectable 
only at extremely high angular resolution. For an extended object, 
the very sharp density peaks smear out depending on the size of the 
object, $r_s$, but the peak densities change significantly only within 
a distance of the order of $r_s$ from the boundaries \citep[see Fig.~6b 
of][]{kwt10}. In order to see the effects of the object size on the 
global structure of the density wake, we compare the density enhancement 
$\alpha$ in Figures~\ref{fig:ptm}, \ref{fig:rss}, and \ref{fig:rsl} for the 
cases of $r_s/r_p=0$, 0.01, and 0.1, respectively, where the object size $r_s$ 
is defined as the gravitational softening radius of a Plummer-type object. 
From Figures~\ref{fig:ptm}a--b, \ref{fig:rss}a--b, and \ref{fig:rsl}a--b, 
one finds that the sharpness at boundaries are reduced with the softened 
gravitational potential, while there are negligible changes in the global 
shape and morphology of the wake. The profiles of $\alpha$ along the spiral 
pattern in Figure~\ref{fig:rss}c show the smoothing of arm boundaries 
compared to Figure~\ref{fig:ptm}c, chiefly by reducing the peak densities. 
Figure~\ref{fig:rsl}c indicates, however, that the minimum value of $\alpha$ 
is also changed in the case of the further extended object with size $r_s$ 
comparable to or larger than half of the interarm spacing for the point mass 
counterpart, reducing the contrast between arm and interarm densities.

\subsection{Velocity}\label{sec:vel}

\begin{figure}
  \plotone{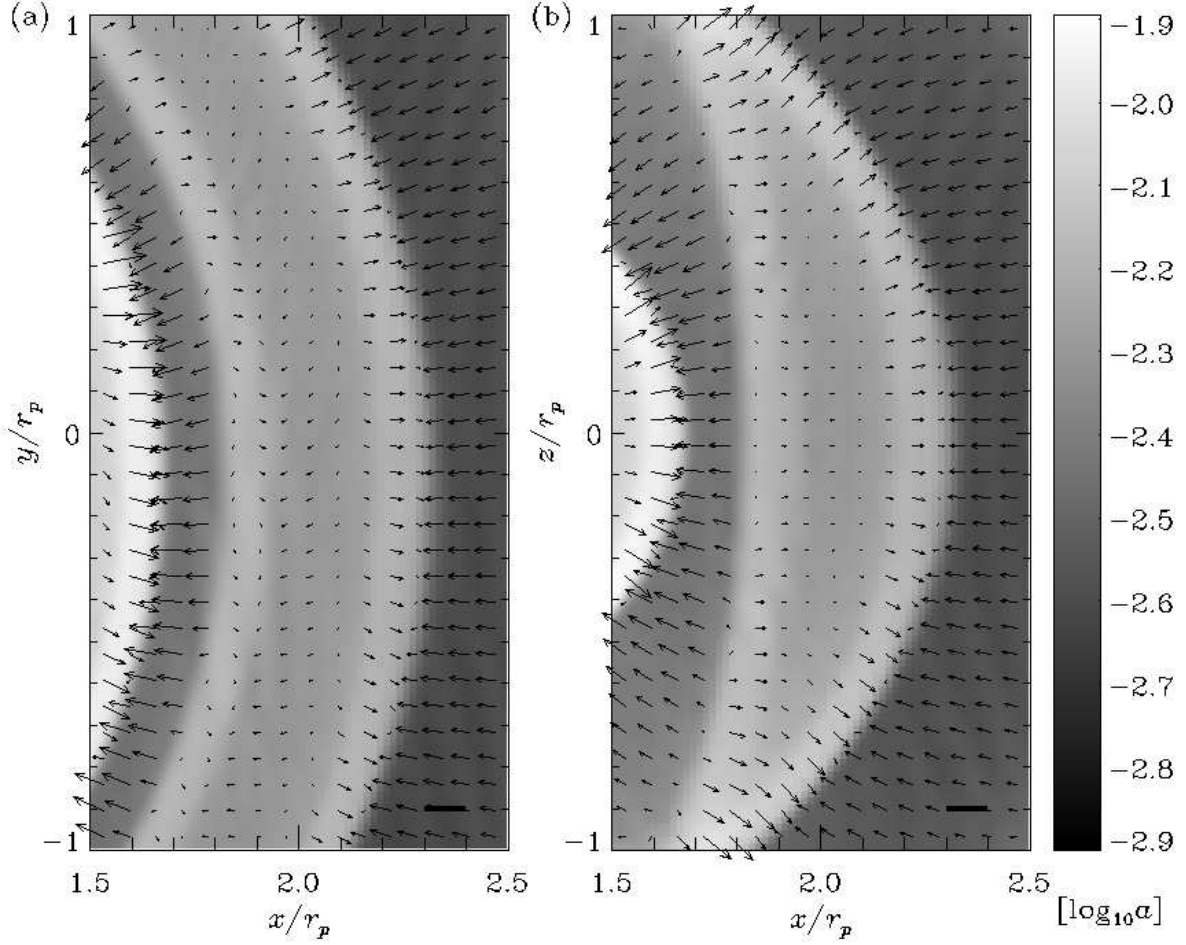}
  \caption{\label{fig:vel}
    Velocity field (arrows) superposed on the map of density enhancement 
    $\alpha$ (logarithmic gray scale) for a segment of high density structure 
    in Fig.~\ref{fig:rss}, observed (a) in the orbital plane ($z=0$) and 
    (b) in a meridional plane ($y=0$). 
    The maximum velocity in each panel approximately corresponds to 
    the line segment at the bottom right corner with the magnitude of
    $\mathcal{M}=r_B\mach^{-1}|r-r_p|^{-1} \times 2\,(r_s/r_p)^{-1/3}$
    in Mach number, where $r_B$, $\mach$, $r_p$, and $r_s$ represent 
    the Bondi accretion radius, orbital Mach number, orbital radius, 
    and gravitational size of the perturbing object, respectively, and 
    $r\sim2r_p$ is adopted.
  }
\end{figure}

In Figure~\ref{fig:vel} zooming in on a high density structure of the wake, 
the initially static medium is revealed to attain an approximately spherical
velocity field. In order to understand the velocity features of the wakes 
for circularly orbiting objects, we first inspect the linear trajectory 
counterparts formulated in \citeauthor{kim09}'s \citeyearpar{kim09} 
equations (A19)--(A20). From their formulae for the velocity components 
$V_\|$ and $V_\bot$ parallel and perpendicular to the object motion, 
respectively, it is straightforward to show that in the linear Mach cone 
(half-opening angle of $\Theta_M=\sin^{-1}\mach^{-1}$) (1) the perturbed 
gas flows toward the edge of the cone perpendicularly, i.e., the fluid 
angle is $\tan^{-1}(|V_\bot/V_\||)\simeq 90\degree-\Theta_M$ in the vicinity 
of the boundary, deflected from $0\degree$ heading for the object behind it 
(see also \citealp[Fig.~1 in][]{san99} and \citealp[Fig.~8 in][]{nam11}); 
and (2) the fluid speed in terms of Mach number approaches the value of the 
density enhancement ($\mathcal{M}\sim\alpha$) close to the boundaries but 
decreases to $\mathcal{M}=r_B\,(\mach\,d)^{-1}$ along the line of the object 
motion at distance $d$ from the object \citep[see also Fig.~1 in][]{kim09}. 
Overall, the fluid in a Mach cone has the maximum velocity near, and toward, 
the boundaries, although it deflects in angle for considerably slower fluid 
parts giving smaller contribution on average.

In the extension of this tendency of the flow to aggregate perpendicularly 
to the shocked boundary, it can be understood that the velocity vectors in 
Figure~\ref{fig:vel} appear to be spherical especially around the tightly 
wound spiral- and circular arc-shaped structures. The spiral boundaries in 
the orbital plane have a small pitch angle\footnote{\ The pitch angle of 
a spiral is defined by $\tan^{-1}\,(\frac{1}{r}\frac{dr}{d\varphi})$, 
which at $r=r_p$ corresponds to $\Theta$ defined in \S\ref{sec:spi}.} 
of $\tan^{-1}(\mach\,r/r_p)^{-1}$ (see \S\ref{sec:spi}) $\le10\degree$ at 
$r/r_p\le3$ for $\mach\ge2$ up to at $r/r_p\le6$ for the extreme case of 
$\mach=1$, making the velocity vectors further spherical at larger distances 
even in the marginally supersonic case. The velocity fields normal to the 
circular arcs in the meridional planes are obviously spherical at large 
distances, notwithstanding that the centers of the arcs are technically 
not at the system center but at $(r,z)/r_p=(\pm1,0)$ (\S\ref{sec:arc}). 
As implied by the above inspection of the velocity field in a Mach cone, 
the flow directions inside the high density structure are not normal to 
the boundary shapes. However, the magnitudes of speeds are negligible 
compared to the maximum speeds around the boundaries, 
$\mathcal{M}=r_B\mach^{-1}|r-r_p|^{-1}\times2\,(r_s/r_p)^{-1/3}$ in 
Mach number, which is empirically found in the range of $\mach=1$--10 
and $r_s/r_p=0.01$--0.1 with $r_B/r_p=0.01$. It also deserves mentioning 
that the flow aggregates toward the outer boundary, while it tends to 
diverge from the inner boundary originated from the behavior of flow 
on a sonic sphere.

\section{NONLINEAR WAKE}\label{sec:non}

The density enhancement of a wake is related to $r_B/r_p$. As this 
value increases, for a fixed $\mach$, the arm density increases and 
finally becomes saturated, showing unexpected behaviors. In previous
numerical simulations \citep{kim09,kwt10}, the nonlinear features 
such as a detached bow shock, turbulence, a disappearance of the 
inner arm boundary, and an increase of background density were 
exhibited without a clear explanation. In this section, we describe 
the nonlinear features of the wake based on the linear analysis in 
the previous section, emphasizing a new explanation for the detached 
bow shock as a phase shift of the inner boundary (defined for the 
corresponding linear wake). With this new explanation, we show that 
the nonlinear features are natural consequences, thus understandable 
in terms of an extension of the properties of linear wakes.

\begin{figure}
  \plotone{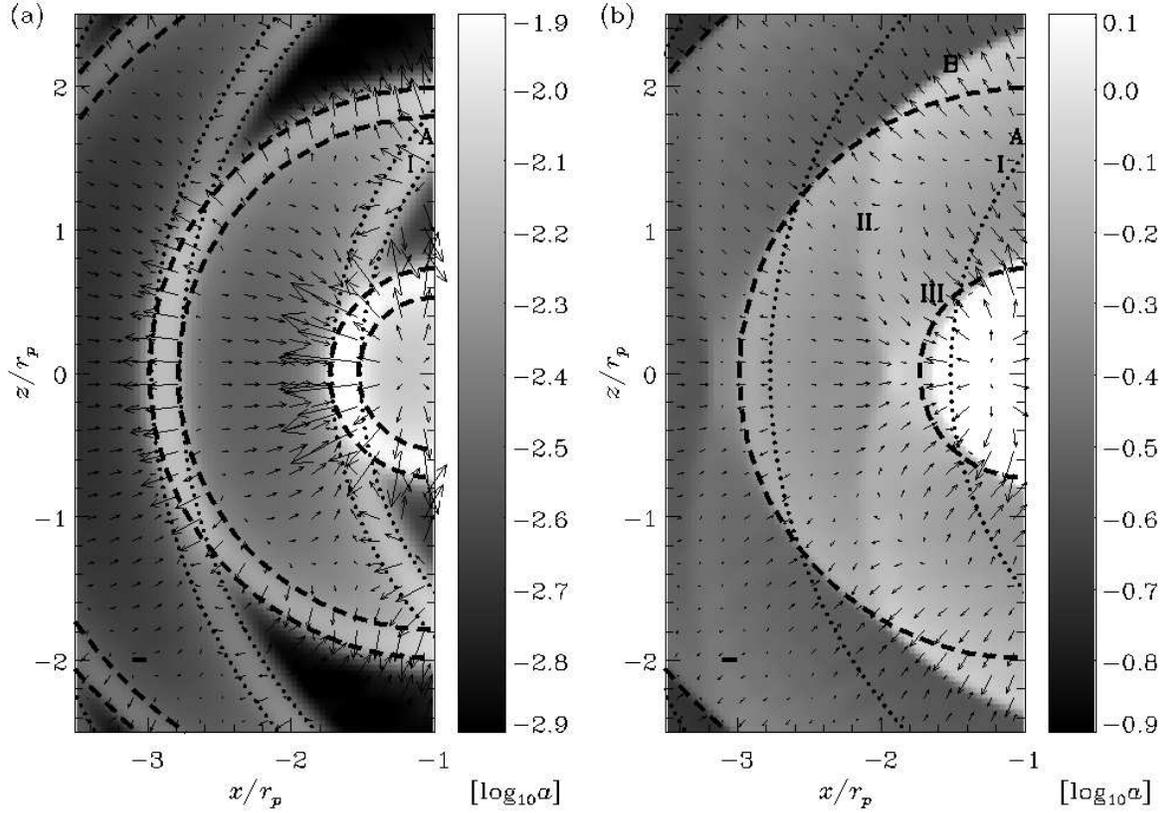}
  \caption{\label{fig:nlv}
    Vertical distributions of density enhancement $\alpha$ (logarithmic gray
    scale) and velocity field (arrows) for the wake created by a gravitational
    potential characterized by its Bondi accretion radius $r_B$ and softening
    radius $r_s=0.1r_p$, where $r_p$ is the orbital radius of the gravitational
    object about origin in $z=0$ plane. The object moving with Mach number 
    $\mach=5$ is currently located at $(x,y,z)/r_p=(1,0,0)$. (a) Under a weak 
    gravitational potential ($r_B/r_p=0.01$), the induced wake has the shape 
    of crescents bounded by circular arcs of radii $4\pi\mach^{-1}+r_s$,
    $6\pi\mach^{-1}+r_s$, and $8\pi\mach^{-1}+r_s$ about the object position 
    with broadening by $\pm r_s$ so as to appear banded ({\it dotted}) and of 
    radii $1\pi\mach^{-1}$, $3\pi\mach^{-1}$, and $5\pi\mach^{-1}$ about the 
    mirror point from the object with the same broadening ({\it dashed}). The 
    material flows approximately from inner dotted lines to outer dashed lines.
    (b) A strong gravitational potential ($r_B/r_p=1.0$) introduces nonlinear
    features of the induced wake, different from the corresponding linear wake 
    in (a). For comparison, the inner dotted and outer dashed lines in (a)
    are copied. Both density and flow speed of nonlinear wake are monotonous
    compared to those of linear wake. The line segment at bottom left corner 
    represents the fiducial magnitude of velocity vectors in Mach number,
    $\mathcal{M}=r_B\mach^{-1}|r-r_p|^{-1}\times2\,(r_s/r_p)^{-1/3}$ with
    adopting $r\sim2r_p$. The spatial resolution is $0.04r_p$ in this area,
    although the maximum resolution achieved around the object position is 
    twice as high with 5 levels of refinement in FLASH code.
  }
\end{figure}

In Figure~\ref{fig:nlv}, we show the vertical structures of wake density 
and velocity field for $\mach=5$, just above a critical Mach number 4.6, 
causing the shallow overlaps between the crescent-shaped overdense regions 
predicted in the linear analysis. The wake shape in Figure~\ref{fig:nlv}a 
(for $r_B/r_p=0.01$) is well explained by the linear prediction in 
\S\ref{sec:arc} (dashed and dotted lines), except for the broadening 
of the edges to form high density belts of width $2r_s$. 
In addition, the inner boundaries of the crescents (dotted lines) are 
enlarged by $r_s$ in radius compared to the prediction, since they form 
when the sonic sphere encounters the curved Mach cone broadened by $r_s$. 
The velocity field in Figure~\ref{fig:nlv}a shows the flow from the inner 
dotted arcs to the outer dashed arcs. These dotted and dashed lines are 
duplicated in Figure~\ref{fig:nlv}b as a reference to the flow pattern 
in the linear wake.

The wake in Figure~\ref{fig:nlv}b is induced by a 100 times stronger 
gravitational potential than in Figure~\ref{fig:nlv}a. This nonlinear 
wake preserves the density discontinuities at the outermost edges of 
the crescent shapes in the linear counterpart (dashed lines), however
they no longer exhibit a belt-like appearance. The expected inner edges 
of crescents (dotted lines) are absent in the nonlinear wake, which are 
in fact shifted outwards beyond the outer edges. For instance, the high 
density band in Figure~\ref{fig:nlv}a near $(x,z)/r_s=(-1.0,1.5)$ (denoted 
by ``A'') tends to fill out the low density region at smaller $z/r_p$ 
as well as to supply additional material to the outer boundary band at 
larger $z/r_p$, achieved only when the perturbation is sufficiently strong 
(Fig.~\ref{fig:nlv}b). The thermal gas pressure close to the crescent's 
apex (``A'') is sufficient in nonlinear regime (Fig.~\ref{fig:nlv}b) to 
spread the material to the point ``B'' beyond the expected outer boundary. 
But the flow originated at a lower latitude (``I'') is unable to reach the 
outer edge of the crescent because of the large distance. Instead, it turns 
back to the inner part, shedding material on the way (``II''), and eventually 
reaches the edge of the smaller crescent defined in linear regime (``III''). 
This process raises the density in between the two shock fronts, forming 
a hill shape of density discontinuity (passing through the point ``II''). 

\begin{figure}
  \plotone{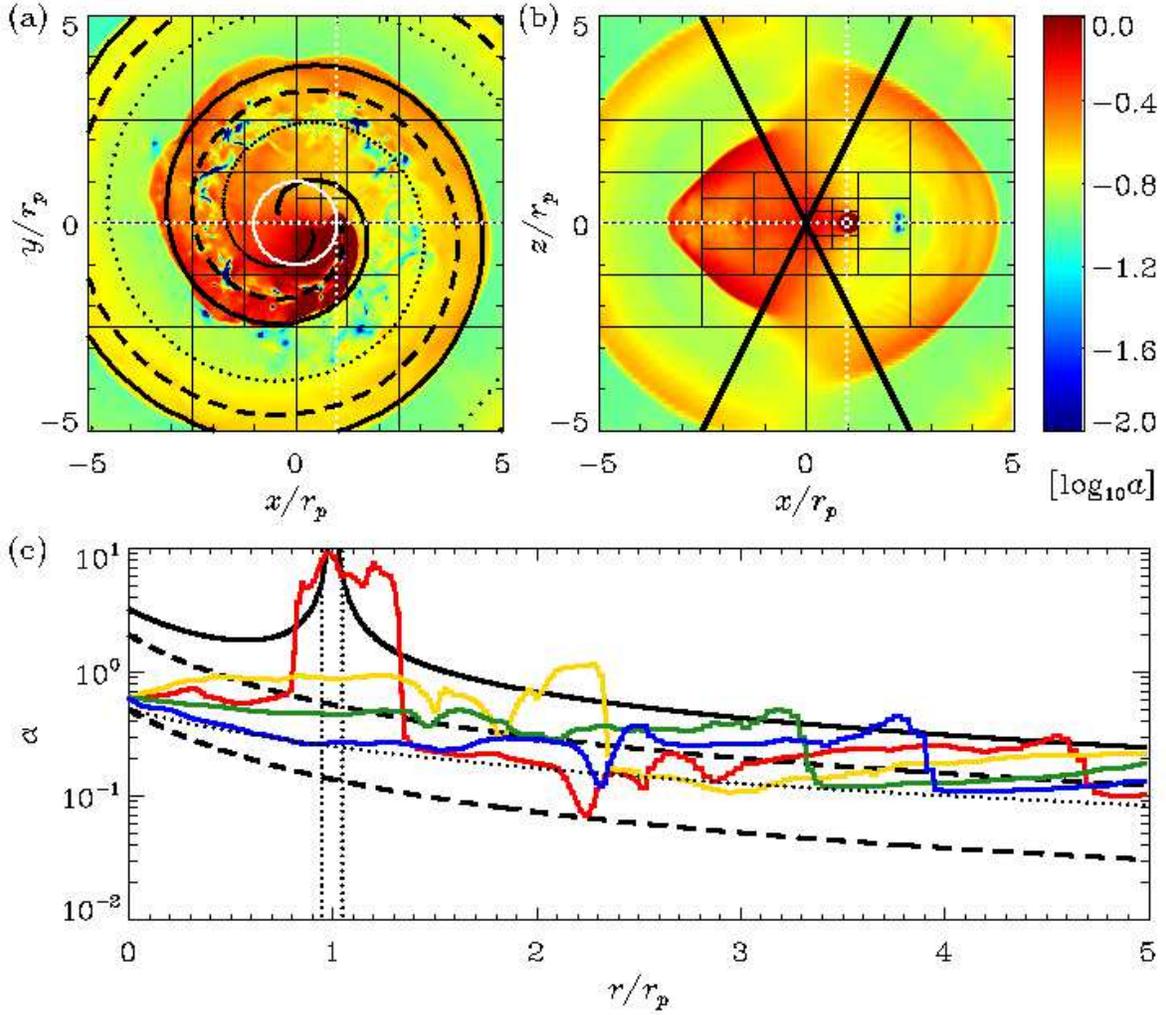}
  \caption{\label{fig:non}
    Density enhancement $\alpha$ of a nonlinear wake due to a gravitational
    potential, characterized by the Bondi accretion radius $r_B=0.5r_p$ and
    the softening radius $r_s=0.05r_p$, in circular motion at distance $r_p$ 
    with the Mach number $\mach=2.2$. The calculation is performed by FLASH 
    code with the maximum refinement level 5 corresponding to a spatial 
    resolution of $0.01r_p$. (a) In the orbital plane, the nonlinear wake 
    departs in shape from the corresponding linear wake of a single-armed 
    spiral confined in the dashed and dotted curves, but its shock front 
    follows the solid curve, which is the dotted line phase shifted by $\pi$. 
    (b) In a meridional plane, the nonlinear wake exhibits the filled 
    elliptical arcs below the solid line as the vertical extension limit 
    of the linear counterpart. (c) The density enhancement profiles along 
    the distance from the center in the orbital plane (colored lines) have 
    the maxima outlined by the black solid line ($\alpha=\alpha_1+6\amin$)
    similarly to the linear counterparts, but the minimum density enhancement 
    follows the dotted line ($\alpha=r_B(r+r_p)^{-1}$) instead of the dashed 
    lines ($\alpha=\amin$ and $4\amin$) for the minimum profiles in linear 
    regime. See text and Fig.~\ref{fig:ptm} for more details.
  }
\end{figure}

An extreme nonlinearity is exhibited in Figure~\ref{fig:non} for the case
of a slower orbital motion ($\mach=2.2$) even with a smaller gravitational 
impact ($r_B/r_p=0.5$) than in Figure~\ref{fig:nlv}b. In fact, the degree 
of nonlinearity can be defined as $\eta_B=r_B/r_p\,(\mach^2-1)^{-1}$ for 
the wake driven by a circularly orbiting object at distance $r_p$, and the 
criterion $\eta_B\sim0.1$ for distinct nonlinear behaviors is found by 
\citet{kwt10}\footnote{\ For rectilinear motion of the object, the similar 
nonlinear features are characterized by the definition of the degree of 
nonlinearity with $r_p$ replaced by the gravitational size $r_s$ of the 
object, $\eta_A=r_B/r_s\,(\mach^2-1)^{-1}$, and the criterion for nonlinear 
regime is $\sim2$ \citep{kim09}.}. In Figure~\ref{fig:non}a, the high density 
region in the orbital plane is not enclosed within the dashed and dotted lines 
representing the spiral arm boundaries of the corresponding linear wake. 
Instead, the sharp discontinuity matches the solid line, which is just the 
inner boundary of the linear counterpart (dotted line) shifted in phase by 
$\pi$. This phase shift is mainly caused by the high pressure at the junction 
($r/r_p=\mach^{-1}$) between the expected outer ($r/r_p=-\varphi\mach^{-1}+1$) 
and inner ($r/r_p=\varphi\mach^{-1}+0.2$) boundaries, if the nonlinear 
effect is ignored, forming a sharp tip. The matter at the tip tends to 
spread the high density over a larger region due to the steep pressure 
gradient, unfolding the inner boundary to match the curvature of the 
outer boundary. This phase shift would be saturated to the maximum value 
of $\pi$ in the very nonlinear case, forming a mirror reflecting curve of 
the inner spiral boundary defined before the nonlinearity is manifested. 

The solid curve in Figure~\ref{fig:non}a can be easily mistaken for a phase
shifted outer boundary of a linear wake \citep[e.g., 75\degree\ for $\mach=2$ 
case in Fig.~3f of][]{kwt10} instead of originating from its inner boundary, 
because the shapes are identical between the two spirals. The $\pi$-shifted 
inner boundary of the linear wake corresponds to the outer boundary of the 
linear wake shifted by $\pi+2(\mach^2-1)^{1/2}-\cos^{-1}(2\mach^{-1}-1)$,
which gives 80\degree\ in advance for $\mach=2.2$. Although it is not clearly
seen in the orbital plane, a density discontinuity is found at the location of 
the outer boundary (defined in the linear counterpart) in Figure~\ref{fig:non}b
for the vertical structure, i.e., the arc structure having $x$-intercept at 
$\sim4r_p$ (see also Fig.~\ref{fig:nlv}b). This presence of the expected 
outer counterpart suggests that the origin of the discontinuity along the 
solid line is not the outer boundary but it is likely interpreted as the 
modified inner boundary of the linear wake with the above explanation. As a 
result, the interior of the nonlinear wake is filled up without a definite 
inner boundary, except for a moderate interference near the black solid 
lines in Figure~\ref{fig:non}b (eq.~[\ref{equ:asy}]). The vertical structure 
of the nonlinear wake is more elliptically shaped than circularly shaped.

As the extension of the solid curve in Figure~\ref{fig:non}a, the 
material immediately behind the object also flows over the curved 
Mach cone shape, advancing 6\degree.7 in angle along the orbit. 
This angle of the isothermal wake ($\gamma=1$) excellently matches 
the fit in \citet{kwt10} for the shock standoff distance in adiabatic 
gas ($\gamma=5/3$). It implies that the shock detachment is a common 
nature of nonlinear flows irrespective of adiabatic index $\gamma$ 
as long as the sonic speed is defined as $\cs=(\gamma p/\rho)^{1/2}$. 
In the process of the overflow creating a new shock front in advance, 
the flow is highly unstable between the old and new shock fronts, 
generating low density eddies in extremely nonlinear cases. According 
to \citet{kim09} for linear trajectory cases, the low density eddies 
survive only when the detached shock distance from the object in the 
lateral direction is larger than $r_B\mach^{-2}$. It is checked that 
the lateral distance of the shock in Figure~\ref{fig:non}a is also 
larger than this value, corresponding to $0.1r_p$.

Lastly, Figure~\ref{fig:non}c shows the density enhancement of the 
wake in this extremely nonlinear regime. The peak values do not have 
a systematic change from the linear prediction (solid curve), while the 
minima do not match any of the dashed curves unlike in the linear regime. 
The minima instead increase to the dotted curve, $\alpha=r_B\,(r+r_p)^{-1}$, 
reducing the arm-interarm density contrast in the entire wake. 
The wake due to circular motion attains a hydrodynamic equilibrium state 
about the system center. In contrast, the wake due to rectilinear motion 
has the equilibrium center about the object itself, so that the amplitude 
of gravitational influence is scaled by $r_B/r_s$ relevant to the object 
size $r_s$. The wake of a circularly orbiting object actually tends to 
achieve quasi-equilibrium about both the object and its orbital center. 
However, the former influences the localized area at distance from the 
object of at most $r_B(\mach^2-1)^{-1}$ (if this distance is greater than 
$r_s$, and $\mach>1$; \citealp{kim09,kwt10}), while the latter affects 
the wake properties globally. The central value of density enhancement 
maintains the value of $r_B/r_p$ independent of the nonlinearity, thus 
readily explaining \citeauthor{kwt10}'s \citeyearpar{kwt10} claim that 
the background density along the orbit is proportional to this value as 
numerically found $\alpha\approx0.46(r_B/r_p)^{1.1}$ in equation (10) of 
\citet{kwt10}.

\section{SUMMARY AND DISCUSSION}\label{sec:sum}

The formation of a gravitational wake is a characteristic feature of 
an astronomical object moving through a medium. 
We examine the wake properties in detail and provide the observable 
aspects in a quantitative way, by revisiting the previous semianalytic 
and numerical works (\kk; \citealp{kwt10}) and comparing the features with 
the linear trajectory counterparts discovered in analytic and numerical 
methods \citep{ost99,kim09}. 

\oops{T}he analytic formulae for the density wake in the previous 
work (reviewed in eq.~[\ref{equ:k07}] and [\ref{equ:sol}]) do not 
give a clear \oops{physical} insight of spiral- and/or arc-shapes. 
Hence it is difficult to deduce the properties of the wake relevant 
to the object properties. A further calculation in this paper 
leads to \oops{a} purely analytic solution for the wake shape 
(eq.~[\ref{equ:mor}]), \oops{which allows detailed description 
of the wake shape}. Simple geometrical approaches are adopted to 
provide the physical understanding of such shapes. Density enhancement 
profiles are also investigated. Since the empirical formulae based on 
the analytic analysis \oops{(eq.~[\ref{equ:alm}]--[\ref{equ:alp}])} 
outline the maximum and minimum density enhancement very successfully, 
these can be directly adopted for interpretation of future simulation and 
observational data. Our deeper understanding of the linear wake helps to 
understand \oops{nonlinear wakes shown in previous numerical simulations}. 
Particularly, nonlinear aspects such as the increase of background density, 
the disappearance of inner arm boundary, and the appearance of a detached 
spiral shock are predictable from a close inspection of linear wake 
properties. 

\oops{Considering} an object characterized by Bondi accretion radius $r_B$ 
\oops{and} circularly orbiting at a distance $r_p$ with an orbital Mach 
number $\mach$ in a static uniform gaseous environment\oops{, o}ur new 
findings for linear wakes are as follows:
\begin{enumerate}
\item In the orbital plane of the perturbing object, the linear wake shows 
  excess density in the form of single-armed structure. For supersonic 
  motion, it is confined by distinct boundaries of Archimedes spirals 
  in the form of $r/r_p=\varphi\mach^{-1}+C$ with different constants 
  $C$'s for the outer (always 1) and the inner (depending on $\mach$) 
  boundaries. This expression for the explicit form of the spiral shape 
  indicates the pitch angle of $\tan^{-1}(\mach\,r/r_p)^{-1}$ and the 
  arm spacing of $2\pi r_p\mach^{-1}$, from which we can deduce the 
  orbital parameters, $\mach$ and $r_p$.
\item The vertical structure 
  of the gravitational density wake is quantified for the first time. 
  In meridional planes, the dense regions are bounded by two-centered 
  concentric arcs forming crescent-like structures ($\mach>1$), partially 
  overlapping each other for high orbital Mach number ($\mach>4.6$) cases. 
  The angular size of the arcs are up to $\lesssim2\tan^{-1}(\mach^2-1)^{1/2}$.
\item The density enhancement $\alpha$ at the orbital center is always
  $r_B/r_p$.
\item The density enhancement at the edges of the spiral arm pattern 
  can be approximated to $\alpha_{\rm arm}=\alpha_1+(4n+2)\;\amin$ with 
  an integer number $n$ depending on $\mach$ and some definitions of 
  $\alpha_1$ (eq.~[\ref{equ:al1}]) and $\amin$ (eq.~[\ref{equ:alm}]). The 
  interarm density also increases to $\alpha_{\rm inter}=(3n-2)\;\amin$ 
  for supersonic cases ($n\geq1$) and $\amin$ for subsonic cases ($n=0$).
  Hence the density contrast between arm and interarm regions, 
  $(\rho_{\rm arm}-\rho_{\rm inter})/\rho_{\rm inter}$ $\approx$ 
  $\alpha_{\rm arm}-\alpha_{\rm inter}$ in the linear regime 
  ($\alpha_{\rm arm},\;\alpha_{\rm inter}\ll1$), is $\sim\alpha_1+(n+4)\,\amin$
  for supersonic cases and $\sim\alpha_1+\amin$ for subsonic cases. 
  The numerical values for density contrast are approximately $r_B/r$ 
  at large distance ($r\gg r_p$), i.e., (0.7--2.9)\,$r_B/r$ for a wide 
  range of orbital Mach numbers $\mach=0$--0.9 and 1.5--10.
\item The initially static fluid acquires velocity components in an 
  approximate spherical form, starting from near the inner boundaries 
  of high density regions and finally accumulating at the outer boundaries.
\item The gravitational size of the object does not significantly affect the 
  linear wake features insofar as the size is smaller than the interarm gap.
\end{enumerate}
As the accretion capability ($r_B/r_p$) increases for a marginally supersonic
orbital Mach number ($\mach$), the thermal pressure can be saturated at the 
junction of high density boundaries appearing as a tip in the linear analysis,
so that the inner boundary extends beyond the outer boundary of the pattern. 
The induced wake reveals the following nonlinear behaviors:
\begin{enumerate}
\item An appearance of a detached bow shock and a disappearance of the
  definite inner arm boundary in the highly nonlinear wake are mutually 
  explained by an overpass of the inner boundary beyond the outer boundary. 
  The shock standoff distance depends only on $r_B$ and $\mach$, regardless 
  of the adiabatic index $\gamma$.
\item Turbulent low density eddies are developed in the unstable process 
  of producing the detached shock via the overpass of the inner boundary
  beyond the outer boundary. These eddies survive when the shock distance 
  in the lateral direction is larger than $r_B\mach^{-2}$, consistent to
  the linear trajectory counterpart in the nonlinear regime. 
\item The density enhancement $\alpha=r_B/r_p$ at the orbital center 
  is preserved even in the extremely nonlinear regime, but the interarm 
  density systematically increases proportional to this value, reducing 
  the arm-interarm density contrast.
\end{enumerate}

Using these theoretical diagnostics, we can derive the object mass, velocity, 
and distance from the orbital center with a given sound speed of the medium.
In the case that concentric arcs are observed, the perturbing object is 
expected to be at the center of the circular arcs, potentially giving $r_p$. 
The angular size of the arcs corresponding to $2\tan^{-1}(\mach^2-1)^{1/2}$ 
probes the orbital Mach number $\mach$ of the object of interest, or the 
corresponding velocity $V_p$ with the aid of an estimate of the sound speed 
$\cs$. The object mass $M_p$ (or the accretion radius $r_B=GM_p/\cs^2$) can 
be roughly estimated from the density contrast $\sim r_B/r$ at a distance 
$r$ from the center. High angular resolution observations are required to 
identify the broadening of arm edges ($\sim2r_s$) so as to estimate the 
object size $r_s$. On the other hand, an observation of a spiral structure 
can provide information on $r_p$ and $\mach$ by measurement of arm spacing 
and arm width. An observation of smooth density distribution without sharp 
discontinuity provides an upper limit of the object speed ($\mach<1$). 
Finally, an observation of the arm structure with a missing inner boundary 
suggests the nonlinear wake condition, $r_B/r_p\,(\mach^2-1)^{-1}\gtrsim0.1$. 

From observations of spirals and/or arcs around stars, one may search 
for the possible existence of unseen companions, which are difficult 
to be directly detected either due to high obscuration by the dense 
environment or due to the bright glare of the central object. 
The presence of the companions may be inferred by their imprints 
on the environment at large distances from the star. For example, 
spiral structures have been observed around AB Aur revealed in 
near-infrared scattered emission \citep{fuk04}, around AFGL 3068 
in dust scattered galactic light \citep{mau06}, and around CIT 6 
in molecular line emission \citep{din09}.

According to the result in this paper, the density contrast of 
a gravitational density wake due to a 10 Jupiter mass planet, 
\oops{orbiting at a distance of several AU}, is 
expected to be of the order of 10\,\% at a distance of a few 
hundreds AU from the central star, where the spiral arms are 
located in the circumstellar disk of the Herbig Ae star AB Aur 
\citep{fuk04}. Although the density contrast of the order of 
10\,\% is not significant, it may be distinguishable in this 
disk (rather than spherical) system. Whether such structures 
can be observed for low mass objects remains to be ascertained. 
And whether the density contrast due to the gravitational influence 
of planets has a similar order of magnitude in a disk system also 
needs to be checked. To clarify the claim of \citet{lin06} of 
forming giant planets as the origin of the observed spiral arms 
around AB Aur, it is necessary to carefully examine the density 
profile of the structure.

In the cases of AFGL 3068 and CIT 6, spiral structures are detected on
the scale of 1--10 thousand AU \citep{mau06,din09}, which exclude the 
possibility of planet wakes since the density contrast decreases with 
distance ($\lesssim1$\,\%). The large spiral structures in these evolved 
circumstellar envelopes are likely to be caused by reflex motion of the 
mass losing star in comparable mass binary system \citep{sok94,mas99,he07,
edg08}. \oops{To date, AFGL 3068 is the only object that shows a clear 
spiral modulation in the circumstellar envelope and its companion star is 
indeed confirmed by a direct imaging in near-infrared \citep{mor06}. The 
binary separation is about a hundred AU.} Recently, \citet{rag11} estimated 
approximately 1 solar mass for the companion of AFGL 3068 from the spiral 
shape. With this companion mass, the gravitational wake is expected to 
contribute as much as 100\,\% of the envelope density at \oops{an observing} 
distance \oops{of the wake} of a thousand AU, doubling the density 
in the arm structure. This density contrast of the companion wake may
be comparable to the density enhancement due to the reflex motion of 
the primary star. Hence, in the central region ($\lesssim$\,1\arcsec\ 
by adopting the distance of 1\,kpc to AFGL 3068), which will be soon 
resolved by the Atacama Large Millimeter/submillimeter Array (ALMA), 
there could exist two separated arms caused by two different mechanisms, 
i.e., one structure due to the reflex motion of the mass losing star 
and one gravitational density wake of the companion star. 
A comparison between the two mechanisms is needed in a range of 
parameter space relevant to the object and background properties.

We note that our analysis is limited to objects in circular motion 
in a static uniform isothermal gaseous medium. For comparison with 
detailed observations, our assumptions for the background medium and 
the object motion may need to be relaxed. However, the wake features 
described in this paper may suffice to provide a zeroth order 
approximation of the primary features expected from the gravitational 
wakes due to such objects.

\acknowledgments

The author is grateful to Ronald E. Taam for continuous discussion 
on this topic through reading the manuscript, and Oscar Morata and 
Woong-Tae Kim for encouraging and helpful comments. The author also 
acknowledges a stimulating report from the anonymous referee, which
increased the significance and potential of this work. \oops{The 
author specially thanks the scientific editor, W. Butler Burton.}
This research is supported by the Theoretical Institute for Advanced 
Research in Astrophysics (TIARA) in the Academia Sinica Institute of 
Astronomy and Astrophysics (ASIAA). 
The numerical simulations presented here are performed using FLASH3.0 code 
developed by the DOE-supported ASC/Alliance Center for Astrophysical 
Thermonuclear Flashes at the University of Chicago.

\clearpage

\end{document}